\documentclass[12pt,aps,pre,preprint,groupedaddress,showpacs,showkeys]{revtex4-1}

\usepackage[utf8]{inputenc}
\usepackage[T1]{fontenc}
\usepackage{graphicx}
\usepackage{url}
\usepackage{hyperref}    
\usepackage{amsmath}

\begin{document}

\title{Monte Carlo simulations of biaxial molecules near a hard wall}



\author{A. Kapanowski}

\email[Corresponding author: ]{andrzej.kapanowski@uj.edu.pl}

\affiliation{Faculty of Physics, Astronomy and Applied Computer Science, 
Jagiellonian University, ulica Łojasiewicza 11, 30-348 Kraków, Poland}

\author{S. Dawidowicz}

\date{\today}

\begin{abstract}
A system of optimal biaxial molecules placed at the sites of a cubic lattice
is studied in an extended Lebwohl-Lasher model. 
Molecules interact only with their nearest neighbors
through the pair potential that depends on the molecule orientations.
It is known that in the homogeneous system there is a direct 
second-order transition from the isotropic to the biaxial nematic phase,
but properties of confined systems are less known.
In the present paper the lattice has periodic boundary conditions 
in the X and Y directions
and it has two walls with planar anchoring, perpendicular to the Z direction.
We have investigated the model using Monte Carlo simulations
on $N_x \times N_y \times N_z$ lattices, $N_x = N_y = 10, 16$, 
$N_z$ from 3 to 19, with and without assuming mirror symmetry.
This study is complementary to the statistical description
of hard spheroplatelets near a hard wall
by Kapanowski and Abram [Phys. Rev. E 89, 062503 (2014)].
The temperature dependence of the order-parameter profiles between walls 
is calculated for many wall separations.
For large wall separations there are the surface layers with biaxial
ordering at both walls (4-5 lattice constants wide)
and beyond the surface layers the order parameters
have values as in the homogeneous system.
For small wall separations the isotropic-biaxial transition is shifted
and the surface layers are thinner.
Above the isotropic-biaxial transition the preferable orientations
in both surface layers can be different.
It is interesting that planar anchoring for biaxial molecules
leads to the uniaxial interactions at the wall.
As a result we get the planar Lebwohl-Lasher model with additional 
(biaxial) interactions with the neighbors from the second layer,
where the Kosterlitz-Thouless transition is present.
\end{abstract}

\pacs{61.30.Cz, 77.84.Nh}

\keywords{liquid crystals, biaxial nematics, Monte Carlo simulations}

\maketitle

\section{Introduction \label{sec:intro}}

Biaxial nematic phases are characterized by an orientational
order along three perpendicular directions $(\vec{L}, \vec{M}, \vec{N})$
and by the existence of three distinct optical axes.
Such phases were first predicted by Freiser in 1970
\cite{1970_Freiser}.
Later, biaxial phases have been studied by mean field theory 
\cite{1974_Straley}, 
\cite{1975_Luckhurst},
\cite{1982_Mulder},
counting methods (a generalization of a Flory's lattice model)
\cite{1972_Shih_Alben},
\cite{1994_Li_Freed},
bifurcation analysis
\cite{1989_Mulder}, and other methods, including computer simulations
\cite{1990_Allen},
\cite{1997_Camp_Allen},
\cite{2020_Skutnik}.
Motivation for these studies ranges from purely academic interest
to the potential usage of biaxial nematics in faster displays.

Straley obtained a phase diagram for a system of biaxial molecules
using mean field theory 
\cite{1974_Straley}.
He showed that four order parameters are necessary to describe
ordered phases with biaxial molecules.
The same was confirmed by Mulder, who derived also the analitical
formula for the excluded volume for a pair of spheroplatelets
which are biaxial objects
\cite{1986_Mulder}.

First theories predict that the system of biaxial molecules can exhibit
four phases, depending on the molecular biaxiality:
the positive uniaxial phase ($N_{U+}$, with prolate molecules),
the negative uniaxial phase ($N_{U-}$, with oblate molecules),
the biaxial phase ($N_B$), and the isotropic phase ($I$).
The nematic-isotropic phase transition is weakly first order
and it becomes continous at the point of maximum molecular biaxiality.
At this point there is a direct transition from the biaxial
to the isotropic phase.

Later theories showed that phase transitions to the biaxial phase can be 
either first or second order with the possibility of several critical points
and reentrant biaxial nematic phases
\cite{2008_Allender}.
In some phase diagrams three different biaxial phases were identified,
where two additional biaxial phases were connected with mixtures
of rodlike and platelike molecules
\cite{2009_Mukherjee}.

\subsection{Lattice models}

The Lebwohl-Lasher (LL) model is a lattice version of the Maier-Saupe model
of anisotropic liquids with uniaxial molecules
\cite{1972_Lebwohl_Lasher},
\cite{1986_Fabbri_Zannoni}.
A weak first-order nematic-isotropic phase transition was found 
in the three-dimensional model at $T^{*}=1.1232(1)$
for lattice sizes up to $28 \times 28 \times 28$
\cite{1992_Zhang}.
Pretransitional fluctuations of the LL model were studied 
by Greeff and Lee
\cite{1994_Greeff_Lee}.
A large lattice of $120 \times 120 \times 120$ was studied on a parallel
supercomputer and the temperature dependence of the energy,
the order parameter and the heat capacity was obtained
with greater accuracy
\cite{1997_Boschi}.
The effect of an external field on a nematic system was also investigated
and the change in the character of the transition from first to second
order with disappearance of the transition at a critical point was observed.

A biaxial version of the LL model was studied by Biscarini \textit{et al.}
\cite{1995_Biscarini}.
They determined the phase diagram of the lattice model
for varying biaxiality. The full set of four second rank order parameters
was calculated for the first time and differences from mean field
theory were discussed.

\subsection{Molecules at the interface}

The properties of the nematic-isotropic phase transition in thin nematic films
were studied for the first time by Sheng \cite{1976_Sheng}.
He used the Landau-de Gennes theory to show the existence
of a critical thickness of the film below which the transition
from the nematic phase to the isotropic phase becomes continuous.
Later this framework was used to describe a boundary-layer phase
transition which occurs at temperatures higher than the bulk-transition 
temperature \cite{1982_Sheng}.

A thin cell with hard spherocylinders was studied by Mao \textit{et al.}
\cite{1997_Mao_Bladon}.
Spherocylinders are composed of cylinders of the length $L$, the diameter $D$,
and hemispherical end caps.
Grand canonical Monte Carlo simulations were used
to investigate the effect of finite aspect ratio $L/D$ 
in density profiles and in order parameters ($L/D = 10, 20$). 
The wall effect penetrated the bulk to a distance of order $L$.
No biaxial order was present in the simulated system if the phase
was isotropic in the bulk.
In the next paper by Mao \textit{et al.} the depletion force was studied
in the confined geometry of two parallel plates
\cite{1997_Mao_Cates}.

In 2000 van Roij \textit{et al.} investigated the phase behavior 
of hard-rod fluid near a single wall and confined in slit pore
\cite{2000_Roij_EPL},
\cite{2000_Roij_JCP}.
They showed a wall-induced surface transition from uniaxial to biaxial
symmetry and complete orientational wetting of the wall-isotropic
fluid interface by a nematic film.
Theoretical analysis was done by employing Zwanzig's rod-model
where the molecules are restricted to orientations
which are parallel to one of the Cartesian coordinate axes.
The results were confirmed by Monte Carlo simulations 
of a fluid of hard spherocylinders with $L/D = 15$
\cite{2001_Dijkstra}.

Liquid crystals confined between parallel walls were studied by Allen
\cite{2000_Allen}.
Computer simulations were compared with the theoretical predictions 
of Onsager’s density-functional theory.
Several different anchoring conditions at the wall-nematic interface
were investigated. In all cases, the principal effect of increasing 
the average density is to increase the surface film thicknes.

A density-functional treatment of a hard Gaussian overlap
fluid confined between two parallel hard walls was presented
by Chrzanowska \textit{et al.} \cite{2001_Chrzanowska}.
For uniaxial particles of elongation 5, 
the density and the order parameter profiles were obtained
in the Onsager approximation.
The surface layers of thickness about half of a particle length
were present with the uniaxial and biaxial order,
in the case of the isotropic and uniaxial phase in the bulk, respectively.

The effect of the incomplete interaction on the nematic-isotropic transition
at the nematic-wall interface was studied by Batalioto \textit{et al.}
\cite{2004_Batalioto}. They used an extended Maier-Saupe approach
with additional interactions with the wall.
In this framework they showed the existence of a boundary layer in which 
the order parameter can be greater or smaller than the one in the bulk, 
according to the strength of the surface potential with respect 
to the nematic one.

The equilibrium phase behavior of a confined rigid-rod system was
studied by Green \textit{et al.} \cite{2010_Green}.
The distribution functions for stable and unstable equilibrium states
were computed as a function of the system density and the system width.
The surprising conclusion was that the introduction of walls perturbs 
the stability limits for any system width, which means that walls always 
impact the interior of systems.

Aliabadi \textit{et al.} examined the ordering properties of \emph{rectangular} 
hard rods at a single planar wall and between two parallel hard walls 
using the second virial density-functional theory
in the Zwanzig approximation \cite{2015_Aliabadi}.
The most interesting finding for the slit pore is the first-order 
transition from the surface ordered isotropic to the capillary nematic phase.
This transition weakens with decreasing pore width and terminates 
in a critical point.

A system of hard spheroplatelets near a hard wall was studied
in the low-density Onsager approximation
by Kapanowski and Abram \cite{PhysRevE.89.062503}.
Spheroplatelets had optimal shape between rods and plates, 
and the direct transition from the isotropic to the biaxial nematic phase 
was present in the bulk.
For the one-particle distribution function $\rho(z,R)$
a simple approximation was used and as a result the order parameters were 
equal to their bulk values unless we were in the interfacial region
thinner then the molecule length.
Biaxiality close to the wall appeared only if the phase was biaxial in the bulk.
For the case of the isotropic phase in the bulk, the phase near the wall
was uniaxial (oblate).

Our aim in the present paper is to get more realistic order parameter profiles
between two walls and check the width of the interfacial region
for the system of biaxial molecules with the direct transition 
from the isotropic to the biaxial nematic phase.
This paper is organized as follows.
The lattice model of biaxial molecules is described in Sec. \ref{sec:system}.
In Sec. \ref{sec:results} we present the results of Monte Carlo simulations
of the homogeneous and confined systems.
Section \ref{sec:conclusions} contains the summary.

\section{System \label{sec:system}}

We have considered a system of optimal biaxial molecules placed at the sites 
of a cubic lattice $N_x \times N_y \times N_z$.
The orientation of a rigid molecule can be determined by several methods:
by the three Euler angles $R = (\phi, \theta, \psi)$,
by the three orthonormal vectors $(\vec{l}, \vec{m}, \vec{n})$,
by the orthogonal rotation matrix,
and by the unit quaternion
\cite{1982_Vesely}.
We are using quaternions in simulations because they are compact,
stable numerically, and we do not have to use slow trigonometric functions.
Our calculations are based on the second rank pair potential
\cite{1975_Luckhurst},
\cite{1995_Biscarini},
\begin{equation}
U(R_{ij}) = - \epsilon_{ij} [ F_{00}^{(2)}(R_{ij}) 
+ \lambda F_{02}^{(2)}(R_{ij})
+ \lambda F_{20}^{(2)}(R_{ij}) 
+ \lambda^2 F_{22}^{(2)}(R_{ij}) ],
\end{equation}
where $R_{ij}$ is the relative orientation of the molecule pair,
$\epsilon_{ij}$ is equal to a positive constant $\epsilon$
for nearest neighbors and zero otherwise.
The biaxiality parameter $\lambda$ accounts for the deviation from cylindrical
molecular symmetry.
For $\lambda=0$ the Lebwohl-Lasher model is recovered.
The value $\lambda=1/\sqrt{3}$ marks the boundary between a system of prolate
($\lambda < 1/\sqrt{3}$) and oblate molecules ($\lambda > 1/\sqrt{3}$).
In our study we focus on the most biaxial molecules so $\lambda=1/\sqrt{3}$.
Note that $\lambda = \sqrt{2} \lambda_B$, where $\lambda_B$ was used
in \cite{1995_Biscarini}; the difference comes from different 
definitions of symmetry-adapted functions.
The functions $F^{(j)}_{\mu\nu}$ are defined in Ref.
\cite{PhysRevE.55.7090}
and they are related to Wigner functions $D^{(j)}_{\mu\nu}$.
The most important functions are

\begin{equation}
F_{00}^{(2)}(R) = P_2(n_z), 
\end{equation}

\begin{equation}
F_{02}^{(2)}(R) = \frac{\sqrt{3}}{3} [P_2(l_z) - P_2(m_z)],
\end{equation}

\begin{equation}
F_{20}^{(2)}(R) = \frac{\sqrt{3}}{3} [P_2(n_x) - P_2(n_y)],
\end{equation}

\begin{equation}
F_{22}^{(2)}(R) = \frac{1}{3} [P_2(l_x) + P_2(m_y) - P_2(m_x) - P_2(l_y)],
\end{equation}
where $P_2(x)$ is the second Legendre polynomial.
For the completely ordered system with all molecule orientations
parallel to the walls we get $U = -3N \epsilon (1+\lambda^2)$,
where $N = N_x N_y N_z$ is the number of molecules (lattice sites).

We have performed Monte Carlo (MC) simulations
with three different boundary conditions.
(i) Periodic boundary conditions in the three directions are for 
the homogeneous system.
(ii) Periodic boundary conditions in the two directions $X, Y$
and two walls at $z=0$ and $z=(N_z-1)a$ with planar anchoring
are for the confined systems, $a$ is the lattice constant.
The distance between walls is $L_z = (N_z-1)a$.
(iii) Periodic boundary conditions in the two directions $X, Y$,
the wall at $z=0$ with planar anchoring, and mirror symmetry applied 
at $z=(N_z-1)a$; this corresponds to $L_z = 2(N_z-1)a$,
but less computer resources are needed for simulations. 
We will compare the results from different simulations when
the conditions (ii) and (iii) describe the same physical system.
Planar anchoring at the walls is motivated by the fact that elongated
molecules can be closer to the wall only if they are parallel to the wall.
On the other hand, the isotropic-nematic interface favors planar
anchoring \cite{2000_Allen}.

Let us move to the determination of the order parameters
and their temperature dependence.
In computer simulations of homogeneous systems three tensors are typically used
\cite{1990_Allen},
\cite{1997_Camp_Allen},
\cite{1995_Biscarini}

\begin{equation}
Q_{\alpha\beta}^{ll} = \frac{3}{2N} \sum_{i=1}^{N} \left(
l_{\alpha}^i l_{\beta}^i - \frac{1}{3} \delta_{\alpha\beta}
\right),
\end{equation}

\begin{equation}
Q_{\alpha\beta}^{mm} = \frac{3}{2N} \sum_{i=1}^{N} \left(
m_{\alpha}^i m_{\beta}^i - \frac{1}{3} \delta_{\alpha\beta}
\right),
\end{equation}

\begin{equation}
Q_{\alpha\beta}^{nn} = \frac{3}{2N} \sum_{i=1}^{N} \left(
n_{\alpha}^i n_{\beta}^i - \frac{1}{3} \delta_{\alpha\beta}
\right).
\end{equation}
Through a diagonalization of these tensors one could determine 
the order parameters according to the procedure described in
\cite{1995_Biscarini}.
The nontrivial problem is finding a consistent way of assigning
the three eigenvalues to the $X,Y,Z$ axes.
In our confined systems all three tensors 
$Q_{\alpha\beta}^{ll}$,
$Q_{\alpha\beta}^{mm}$,
$Q_{\alpha\beta}^{nn}$ 
are calculated independently for all layers parallel to the walls.
Then the $Z$ axis is always perpendicular to the walls and the $X$ axis
is parallel to the walls and corresponds to the maximum eigenvalue
of the tensor $Q_{\alpha\beta}^{nn}$.
Finally, four order parameters $\langle F_{\mu\nu}^{(2)} \rangle$ 
are calculated

\begin{equation}
\langle F_{00}^{(2)} \rangle = Q_{zz}^{nn} = - Q_{zz}^{ll} - Q_{zz}^{mm},
\end{equation}

\begin{equation}
\sqrt{3} \langle F_{02}^{(2)} \rangle = Q_{zz}^{ll} - Q_{zz}^{mm}
= -Q_{xx}^{ll} - Q_{yy}^{ll} + Q_{xx}^{mm} + Q_{yy}^{mm},
\end{equation}

\begin{equation}
\sqrt{3} \langle F_{20}^{(2)} \rangle = Q_{xx}^{nn} - Q_{yy}^{nn}
= -Q_{xx}^{ll} - Q_{xx}^{mm} + Q_{yy}^{ll} + Q_{yy}^{mm},
\end{equation}

\begin{equation}
3 \langle F_{22}^{(2)} \rangle 
= Q_{xx}^{ll} + Q_{yy}^{mm} - Q_{yy}^{ll} - Q_{xx}^{mm}.
\end{equation}
Note that the same order parameters must have the same values in all
the ways they are computed.
The values of the order parameters depend on the phase orientation.
For the completely ordered system there are six main phase orientations
which are sumarized in Table \ref{tab:Fjmn}.
The $\langle F_{00}^{(2)} \rangle$ order parameter is a measure of the alignment 
of the $\vec{n}$ molecule axis along the $Z$ axis of the reference frame. 
The $\langle F_{02}^{(2)} \rangle$ order parameter describes the relative 
distribution of the $\vec{l}$ and the $\vec{m}$ axes along the $Z$ axis. 
Both $\langle F_{00}^{(2)} \rangle$ and $\langle F_{02}^{(2)} \rangle$
can be nonzero in the uniaxial nematic phase. 
The $\langle F_{20}^{(2)} \rangle$ order parameter describes the relative
distribution of the $\vec{n}$ axis along the $X$ and the $Y$ axes. 
The $\langle F_{22}^{(2)} \rangle$ order parameter is
related to the distributions of the $\vec{l}$ and $\vec{m}$ axes 
along the $X$ and the $Y$ axes.
Both $\langle F_{20}^{(2)} \rangle$ and $\langle F_{22}^{(2)} \rangle$
signal biaxiality of the phase.

\begin{table}
\caption{
\label{tab:Fjmn} Order parameters (OP) for the completely ordered systems.
Orientations used for the homogeneous and the confined systems are marked.}
\begin{ruledtabular}
\begin{tabular}{lcccccc}
OP & \shortstack{
$\vec{N}||Z, \vec{L}||X$ \\ (homogeneous)} &
$\vec{N}||Z, \vec{L}||Y$ &
\shortstack{
$\vec{N}||X, \vec{L}||Z$ \\ (confined)} &
$\vec{N}||X, \vec{L}||Y$ &
$\vec{N}||Y, \vec{L}||Z$ &
$\vec{N}||Y, \vec{L}||X$  \\ \hline
$\langle F_{00}^{(2)} \rangle$ & $1$ & $1$ & $-1/2$       & $-1/2$        & $-1/2$        & $-1/2$  \\
$\langle F_{02}^{(2)} \rangle$ & $0$ & $0$ & $\sqrt{3}/2$ & $-\sqrt{3}/2$ & $\sqrt{3}/2$  & $-\sqrt{3}/2$  \\
$\langle F_{20}^{(2)} \rangle$ & $0$ & $0$ & $\sqrt{3}/2$ & $\sqrt{3}/2$  & $-\sqrt{3}/2$ & $-\sqrt{3}/2$  \\
$\langle F_{22}^{(2)} \rangle$ & $1$ & $-1$ & $1/2$       & $-1/2$        & $-1/2$        & $1/2$  \\
\end{tabular}
\end{ruledtabular}
\end{table}

\section{Results \label{sec:results}}

\begin{figure}
\includegraphics{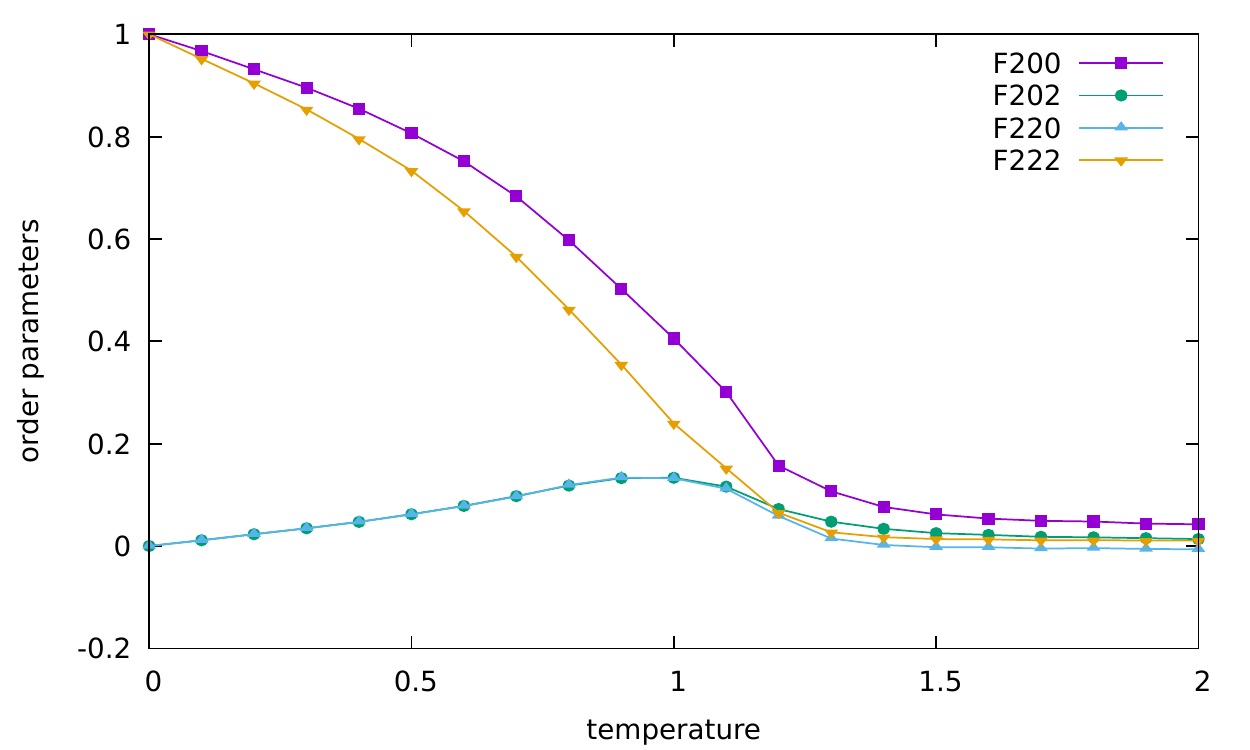}
\caption{
\label{fig:Fjmn_bulk} 
Order parameters $\langle F_{\mu\nu}^{(j)} \rangle$ ($Fj\mu\nu$ in the picture)
vs temperature for the homogeneous system.
Results obtained from $10 \times 10 \times 10$ MC for $\lambda=1/\sqrt{3}$.
The biaxial-isotropic transition is near $T = 1.18 \epsilon/k_B$.}
\end{figure}

Prior to study the confined systems we have calculated the homogeneous
bulk system and the temperature dependence of the order parameters,
as shown in Fig. \ref{fig:Fjmn_bulk}.
We have performed MC simulations on $10 \times 10 \times 10$
and $16 \times 16 \times 16$ lattices with
periodic boundary conditions in all three directions.
The temperature step was typically $0.1$ and $0.01$ near the $I-N_B$ transition.
We have used $10^4$ lattice cycles for warmup and $10^4$ cycles for production,
where a cycle is $N$ attepted moves. 
Sometimes $10^5$ cycles were used as an additional check.
We started from the ideal configuration at the lowest temperature,
then the last configuration for a given temperature was used as the initial
configuration for the next temperature.
Figure~\ref{fig:Fjmn_bulk} shows that the biaxial-isotropic transition
is near $T = 1.18 \epsilon/k_B$, in agreement with
\cite{1995_Biscarini}.
The energy of the homogeneous system is always negative and 
it is an increasing function of temperature.

\begin{figure}
\includegraphics{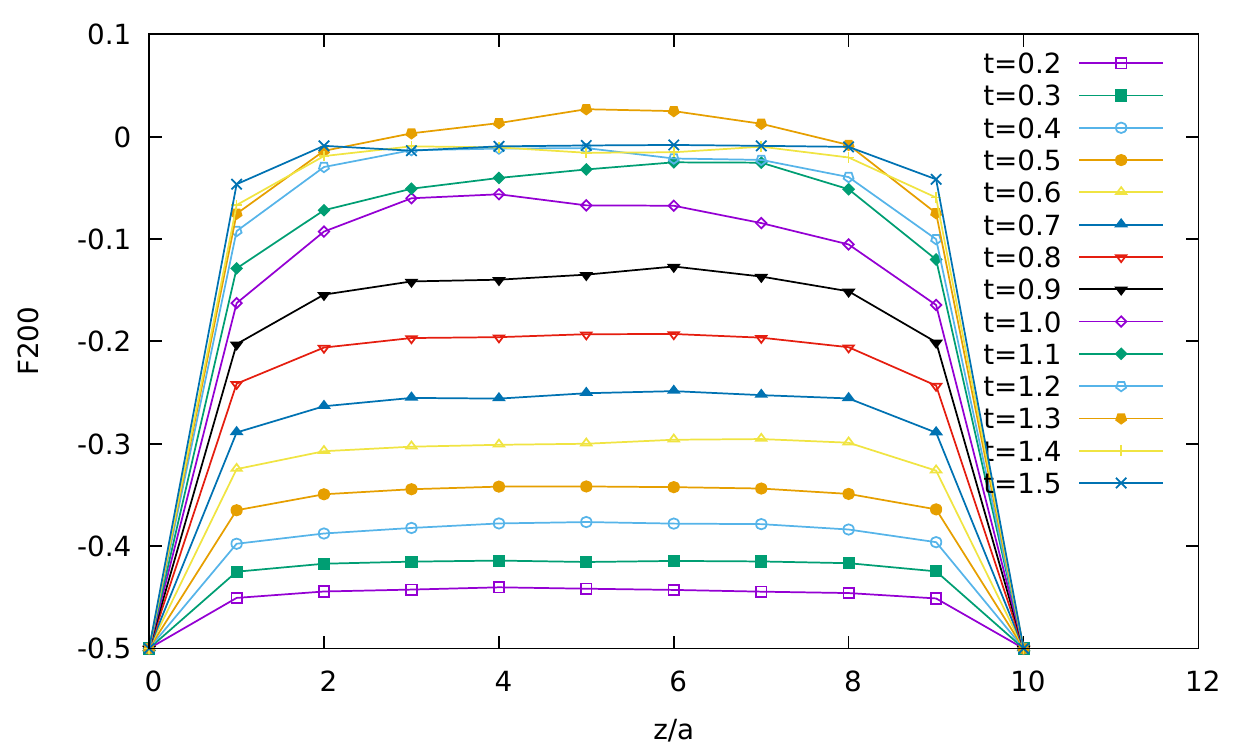}
\caption{
\label{fig:F200_2walls} 
The profiles of $\langle F_{00}^{(2)} \rangle$ vs $z$ 
for the confined system between two walls.
Results obtained from $10 \times 10 \times 11$ MC for $\lambda=1/\sqrt{3}$,
$t = k_B T/\epsilon$.}
\end{figure}

\begin{figure}
\includegraphics{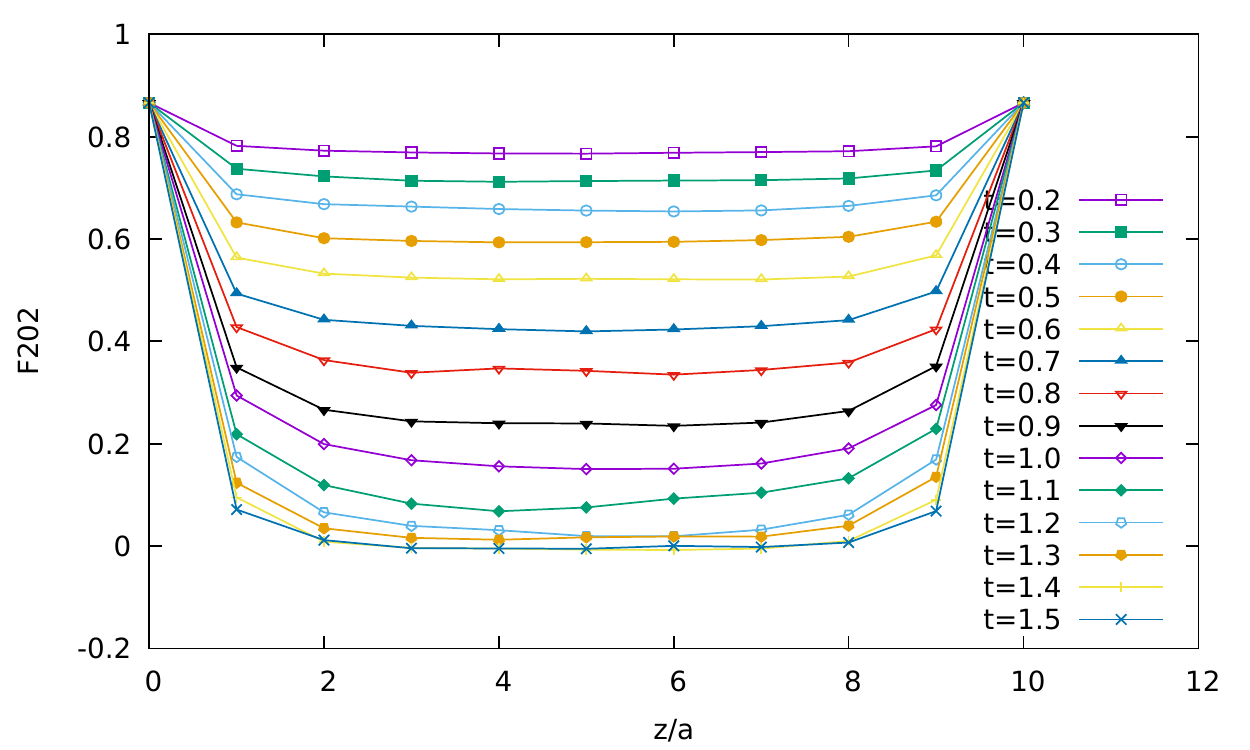}
\caption{
\label{fig:F202_2walls} 
The profiles of $\langle F_{02}^{(2)} \rangle$ vs $z$ 
for the confined system between two walls.
Results obtained from $10 \times 10 \times 11$ MC for $\lambda=1/\sqrt{3}$,
$t = k_B T/\epsilon$.}
\end{figure}

\begin{figure}
\includegraphics{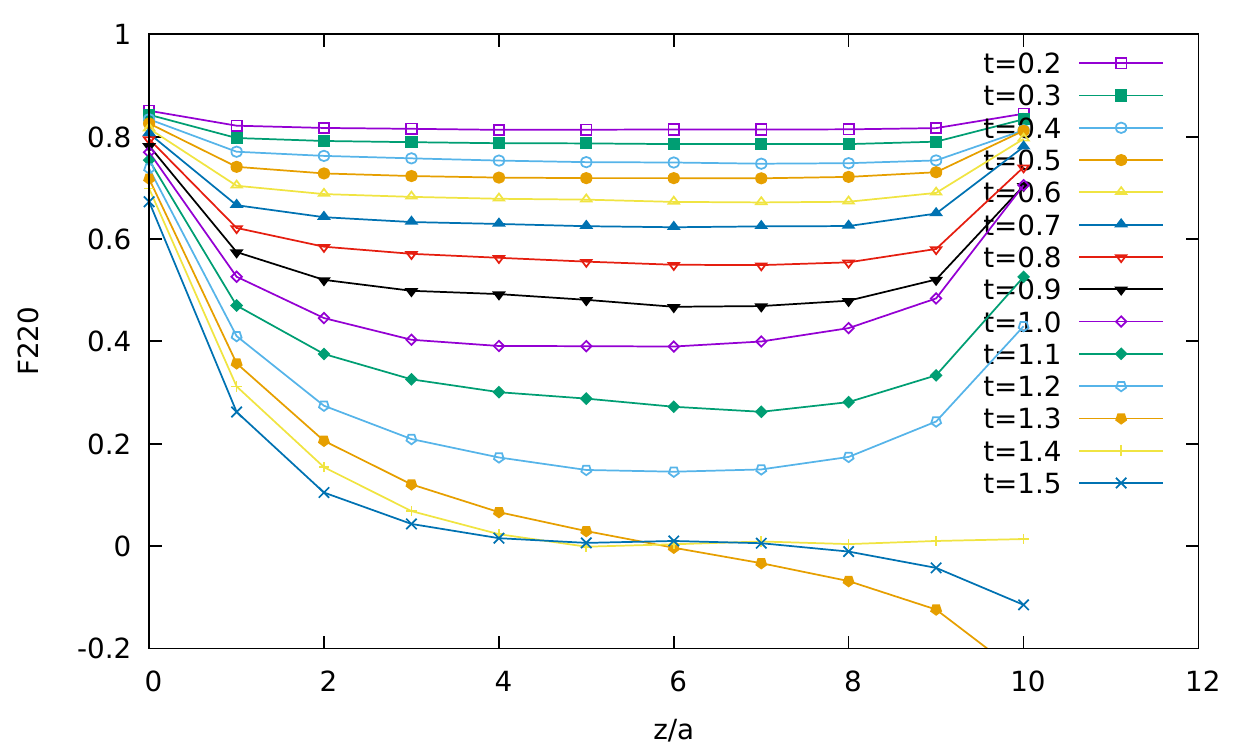}
\caption{
\label{fig:F220_2walls} 
The profiles of $\langle F_{20}^{(2)} \rangle$ vs $z$ 
for the confined system between two walls.
Results obtained from $10 \times 10 \times 11$ MC for $\lambda=1/\sqrt{3}$,
$t = k_B T/\epsilon$.
Different preferable orientations on the walls are visible
for high temperatures.}
\end{figure}

\begin{figure}
\includegraphics{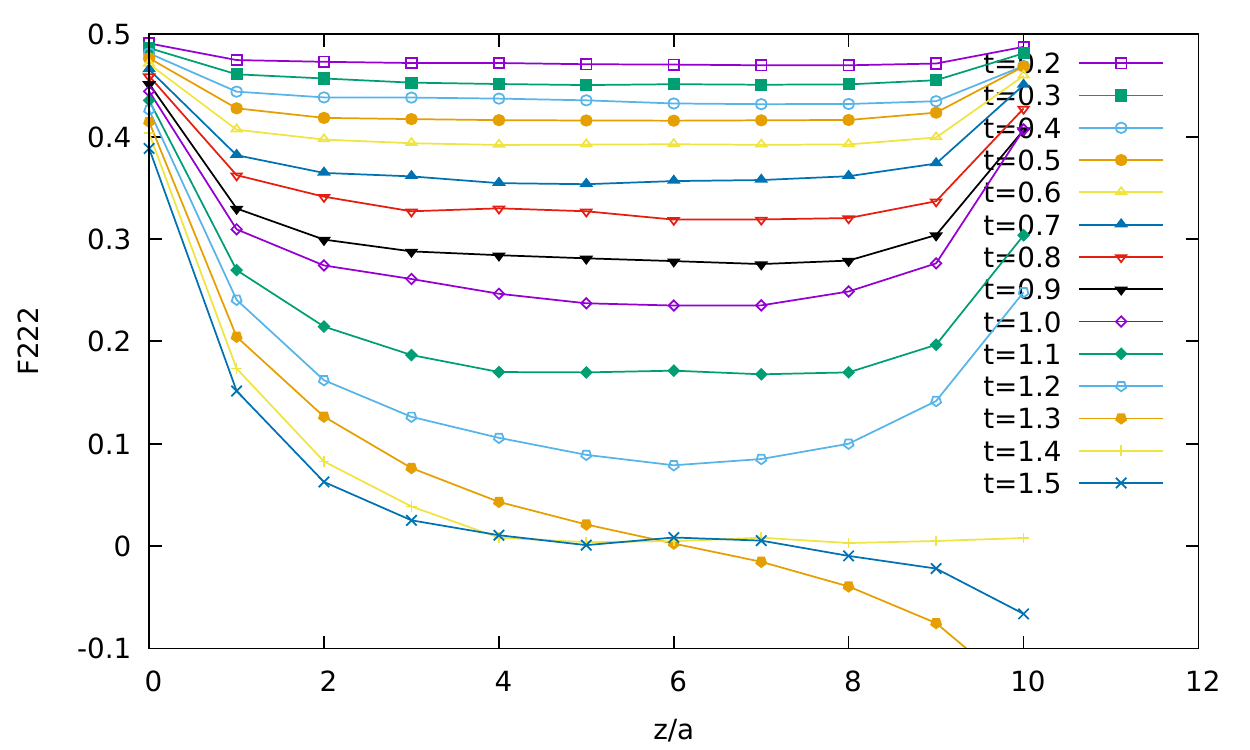}
\caption{
\label{fig:F222_2walls} 
The profiles of $\langle F_{22}^{(2)} \rangle$ vs $z$ 
for the confined system between two walls.
Results obtained from $10 \times 10 \times 11$ MC for $\lambda=1/\sqrt{3}$,
$t = k_B T/\epsilon$. 
Different preferable orientations on the walls are visible
for high temperatures.}
\end{figure}

\begin{figure}
\includegraphics[scale=0.25]{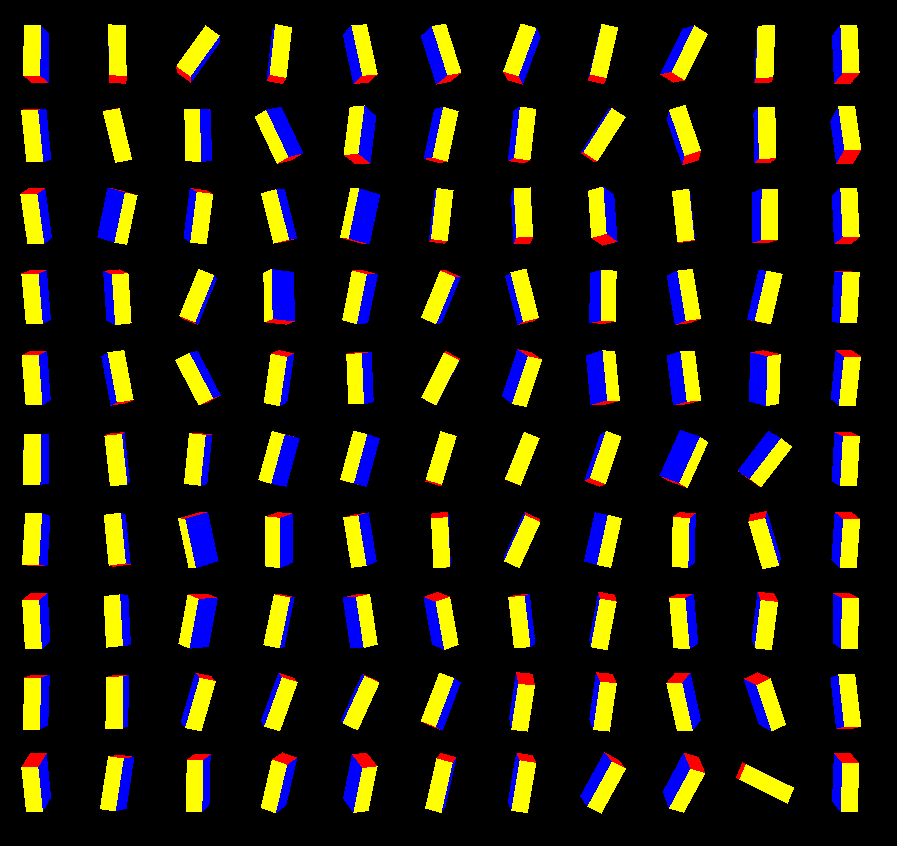}
\caption{
\label{fig:nem_phase} 
A snapshot of simulation configuration ($YZ$ layer)
in the biaxial nematic phase at $T=0.5\epsilon/k_B$
for the confined system between two walls.
Results obtained from $10 \times 10 \times 11$ MC for $\lambda=1/\sqrt{3}$.
Long molecule axes are parallel to the $Y$ axis, 
short molecule axes are parallel to the $Z$ axis.}
\end{figure}

\begin{figure}
\includegraphics[scale=0.25]{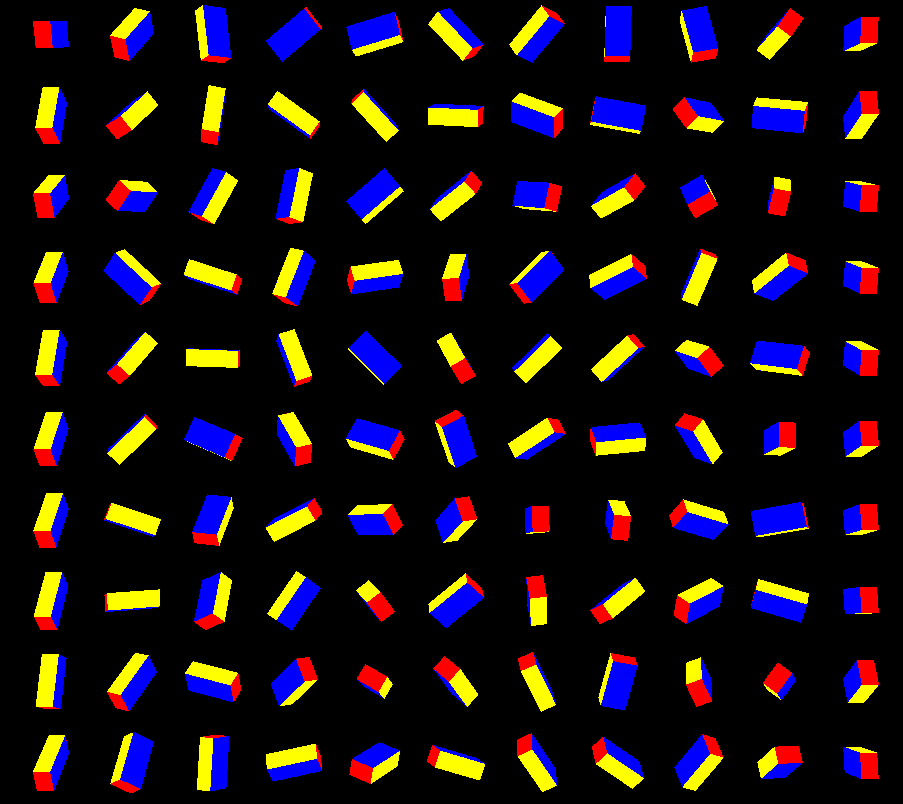}
\caption{
\label{fig:iso_phase} 
A snapshot of simulation configuration ($YZ$ layer)
in the isotropic phase at $T=1.5\epsilon/k_B$
for the confined system between two walls.
Results obtained from $10 \times 10 \times 11$ MC for $\lambda=1/\sqrt{3}$.
The prefered orientations of the molecules on both walls 
(the left and the right columns) are different.}
\end{figure}

Let us move to the description of the confined systems.
We have performed MC simulations on $10 \times 10 \times N_z$
lattices, $N_z$ from 3 to 19, with periodic boundary conditions 
in the $X,Y$ directions and two parallel walls with planar anchoring.
Figures from \ref{fig:F200_2walls} to \ref{fig:F222_2walls}
show the order parameters $\langle F_{\mu\nu}^{(2)} \rangle$ profiles
for the lattice system $10 \times 10 \times 11$.
In the isotropic phase ($T > 1.3\epsilon/k_B$) the order parameters 
are almost zero except in the surface layers of the length approximately 
4-5 lattice constants.
Near the walls long molecule axes are nearly parallel to the walls
and this yields $\langle F_{00}^{(2)} \rangle < 0$,
with the expected limit of $-1/2$ at the walls.
On decreasing temperature, a transition take place to the biaxial
nematic phase, at which all order parameters become finite
beyond the surface layers.
Snapshots of simulation configurations
in the biaxial nematic and in the isotropic phases are given 
in Fig. \ref{fig:nem_phase} and Fig. \ref{fig:iso_phase}, respectively.
In the biaxial nematic phase the preferable orientation of molecules on both
walls is the same, although it is changing during computations.
In the isotropic phase different preferable orientations on the walls
are common and it is visible in the snapshots and the order parameter profiles.
We note that this effect can not be obtained using the boudary conditions
with mirror symmetry. What is more, we have got unphysical effects
in the cell center, probably due to inconsistency in the formula for
the total energy of the system.

\begin{figure}
\includegraphics{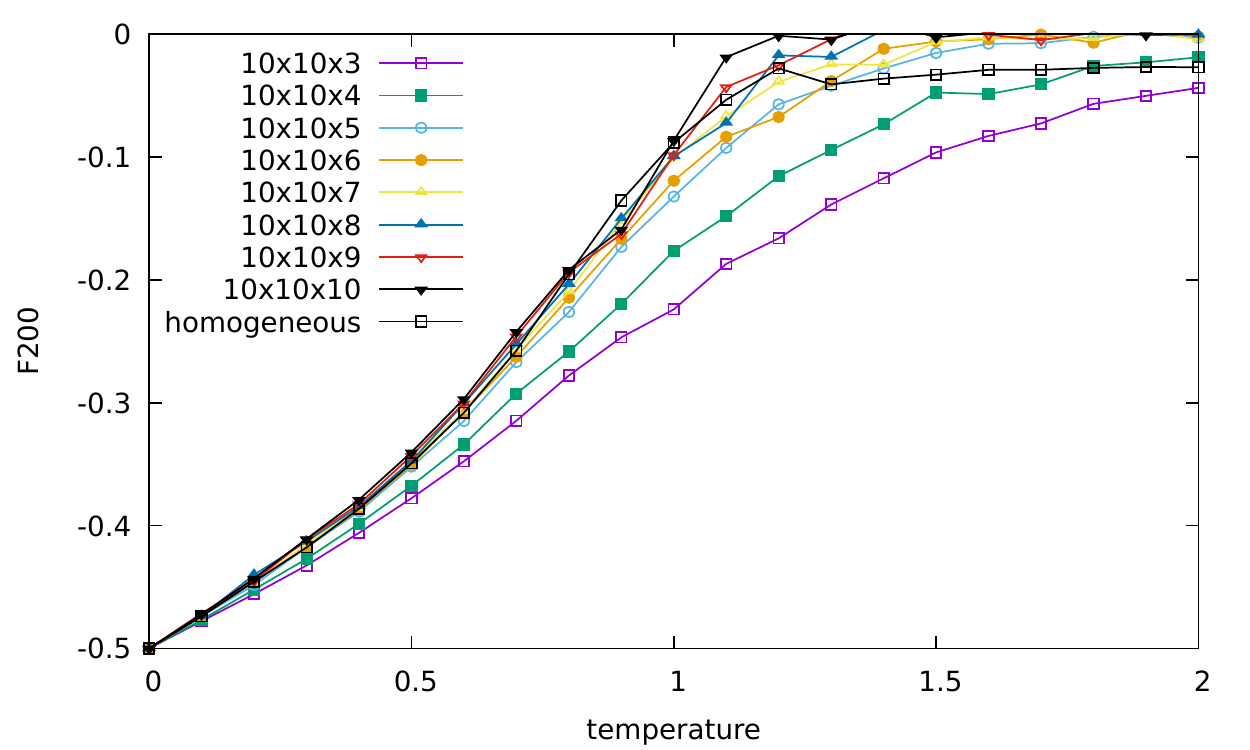}
\caption{
\label{fig:F200_center} 
Temperature dependence of the order parameter $\langle F_{00}^{(2)} \rangle$
for the confined systems between two walls in the cell center.}
\end{figure}

\begin{figure}
\includegraphics{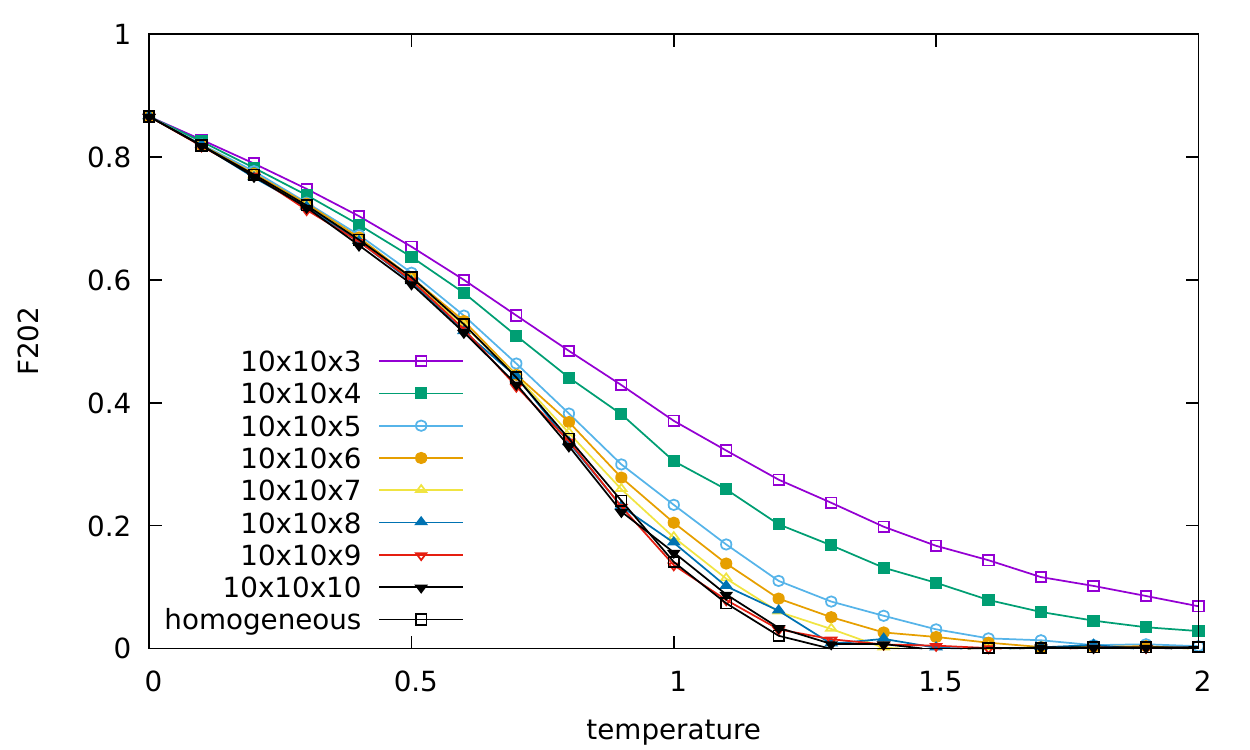}
\caption{
\label{fig:F202_center} 
Temperature dependence of the order parameter $\langle F_{02}^{(2)} \rangle$
for the confined systems between two walls in the cell center.}
\end{figure}

\begin{figure}
\includegraphics{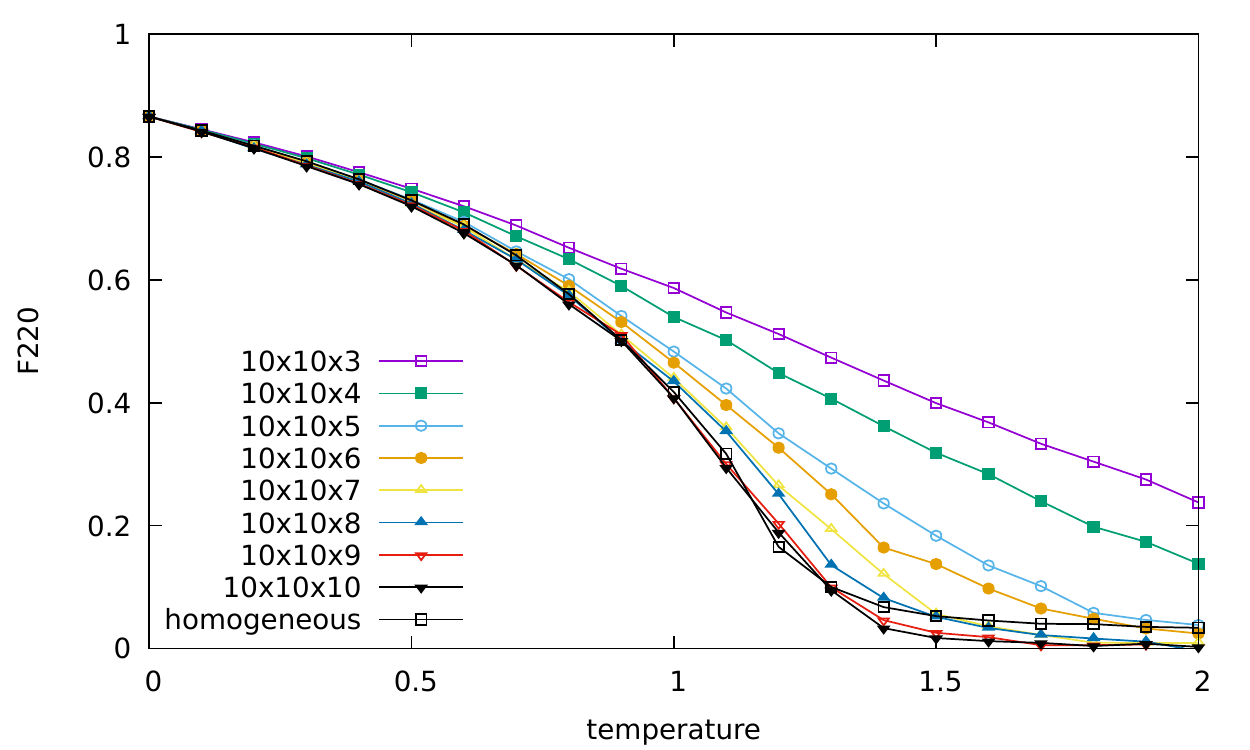}
\caption{
\label{fig:F220_center} 
Temperature dependence of the order parameter $\langle F_{20}^{(2)} \rangle$
for the confined systems between two walls in the cell center.}
\end{figure}

\begin{figure}
\includegraphics{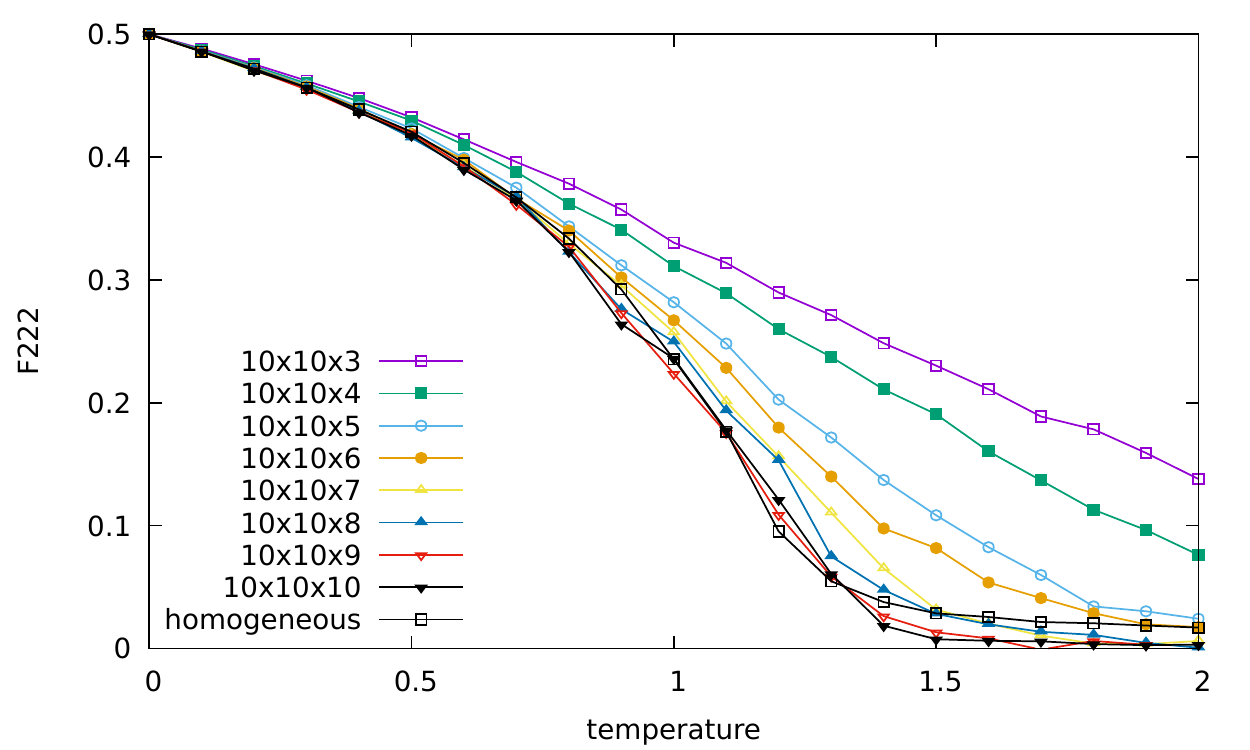}
\caption{
\label{fig:F222_center} 
Temperature dependence of the order parameter $\langle F_{22}^{(2)} \rangle$
for the confined systems between two walls in the cell center.}
\end{figure}

Figures from \ref{fig:F200_center} to \ref{fig:F222_center} show
the temperature dependence of the order parameters in the cell center.
The temperature of the isotropic-biaxial transition is shifted
but for $N_z > 10$ it is almost the same as in the homogeneous system.
From this point the surface layers are separated and they have both
biaxial nematic ordering ($\langle F_{22}^{(2)} \rangle$ is nonzero).
We note a small discrepancy between the results for the homogeneous
system and for the confined systems. This is due to numerical errors
during diagonalization of the tensors $Q_{\alpha\beta}$.
The results for the confined systems are more exact because at the wall
one axis is fixed as perpendicular to the wall.
For $N_z=3$ and $N_z=4$ the biaxial nematic phase is present is the cell
for higher temperatures but the order parameters monotonically
go to zero. We have not found any capilary nematization transition.

\section{Conclusions \label{sec:conclusions}}

In this work we have studied the order-parameter profiles
in the confined systems of optimal biaxial molecules using Monte Carlo 
simulations in an extended Lebwohl-Lasher model.
In the homogeneous system there is a direct second-order 
isotropic-biaxial transition.
We have studied the confined systems with two parallel walls
with planar anchoring and with different wall separations.

For large wall separations there are the surface layers at both walls
with the width of 4-5 lattice constants and beyond the surface layers 
the order parameters have values as in the homogeneous system.
The ordering within the surface layers is always biaxial wheres
in the paper \cite{PhysRevE.89.062503}
biaxiality close to the wall was present only if the phase was biaxial 
in the bulk.
The reason for this discrepancy is planar anchoring at the walls
which creates the planar (uniaxial) Lebwohl-Lasher model with 
the Kosterlitz-Thouless transition
\cite{1988_Chiccoli},
\cite{2003_Mondal}.
In our systems there are additional (biaxial) interactions with neighbors
in the second layer.
The partial ordering at the walls in our finite systems creates
the biaxial ordering in the surface layers for all temperatures.
We note that the surface transition was studied, for uniaxial molecules
and different surface couplings, using the Landau-de Gennes approach
\cite{1993_Lvov},
\cite{1994_Kothekar}.
Additional effects due to external fields were studied in
\cite{2005_Ito} but again for uniaxial molecules.

For small wall separations the isotropic-biaxial transition is shifted
to higher temperatures and the surface layers are thinner.
The preferable orientation of the biaxial nematic phase is approximately 
the same near the walls and in the center of the cell but its direction
can change during simulations.
Above the isotropic-biaxial transition the preferable orientations
in both surface layers can be different.

In summary, the presented results of MS simulations revealed
effects which combine the properties of two-dimensional and
three-dimensional systems. It is important to study systems with
biaxial molecules using different techniqes in order to better
understand the biaxial nematic phases and to find hints for experiments.


\bibliography{kapanowski,biaxial,lattice,wall}

\end{document}